\documentclass[aps,prb,superscriptaddress,twocolumn, showpacs]{revtex4}
\usepackage{amssymb}
\usepackage{graphicx}

\begin{document}
\title{Magnetic, transport, and thermal properties of single crystals of the layered arsenide BaMn$_2$As$_2$}
\author{Yogesh Singh}
\affiliation{Ames Laboratory and Department of Physics and Astronomy, Iowa State University, Ames, IA 50011}
\author{A. Ellern}
\affiliation{Department of Chemistry, Iowa State University, Ames, IA 50011}
\author{D. C. Johnston}
\affiliation{Ames Laboratory and Department of Physics and Astronomy, Iowa State University, Ames, IA 50011}
\date{\today}

\begin{abstract}
Growth of BaMn$_2$As$_2$ crystals using both MnAs and Sn fluxes is reported.  Room temperature crystallography, anisotropic isothermal magnetization $M$ versus field $H$ and magnetic susceptibility $\chi$ versus temperature $T$, electrical resistivity in the $ab$ plane $\rho(T)$, and heat capacity $C(T)$ measurements on the crystals were carried
out.  The tetragonal ThCr$_2$Si$_2$-type structure of BaMn$_2$As$_2$ is confirmed.  After correction for traces of ferromagnetic MnAs impurity phase using $M(H)$ isotherms, the inferred intrinsic $\chi(T)$ data of the crystals are anisotropic with $\chi_{ab}/\chi_{c} \approx 7.5$ at $T = 2$~K\@.  The temperature dependences of the anisotropic $\chi$ data suggest that BaMn$_2$As$_2$ is a collinear antiferromagnet at room temperature with the easy axis along the $c$ axis, and with an extrapolated N\'eel temperature $T_{\rm N} \sim 500$~K\@.  The $\rho(T)$ decreases with decreasing $T$ below 310~K but then increases  below $\sim50$~K, suggesting that BaMn$_2$As$_2$ is a small band-gap semiconductor with an activation energy of order 0.03~eV\@.  The $C(T)$ data from 2 to 5~K are consistent with this insulating ground state, exhibiting a low temperature Sommerfeld coefficient $\gamma = 0.0(4)$~mJ/mol~K$^2$.  The Debye temperature is determined from these data to be $\theta_{\rm D} = 246(4)$~K\@.  BaMn$_2$As$_2$ is a potential parent compound for ThCr$_2$Si$_2$-type superconductors. 

\end{abstract}
\pacs{74.10.+v, 75.30.-m, 75.30.Gw, 75.40.Cx}

\maketitle

\section{Introduction}
\label{sec:INTRO}
The family of compounds $A$Fe$_2$As$_2$ ($A$~=~Ba, Sr, Ca, and Eu) crystallize in the tetragonal ThCr$_2$Si$_2$ type structure and has FeAs layers and $A$ layers alternately stacked along the $c$-axis.  These materials show antiferromagnetic (AF) and structural transitions at high temperatures.\cite{Rotter2008,Krellner2008,Ni2008,Yan2008,Ni2008a,Ronning2008,Goldman2008,Tagel2008,Ren32008,Jeevan2008}  When the $A$ atoms are partially replaced by K, Na, or Cs, the AF and structural transitions are suppressed and superconductivity is observed.\cite{2Rotter2008,2GFChen2008,2Jeevan2008,Sasmal2008}  Even in-plane doping by partially replacing Fe by Co\cite{Sefat2008, Jasper2008} or Ni\cite{Li2008} leads to superconductivity.  
It is of interest to look for other materials with related structures and investigate their physical properties to see if these can be potential parent compounds for new high temperature superconductors.  The undoped Ni-based materials BaNi$_2$As$_2$ (Ref.~\onlinecite{Ronning2008a}) and SrNi$_2$As$_2$ [Ref.~\onlinecite{Bauer2008}] are themselves low temperature superconductors whereas BaCo$_2$As$_2$ is a correlated metal situated near a ferromagnetic instability.\cite{Sefat2008a}  
The isostructural compound BaMn$_2$As$_2$ was previously synthesized in polycrystalline form and its ThCr$_2$Si$_2$-type crystal structure was reported.\cite{Brechtel1978}  To the best of our knowledge the physical properties of BaMn$_2$As$_2$ have not been investigated before.

Herein we report the growth, single crystal structure, electrical resistivity $\rho$ in the $ab$ plane versus temperature $T$, magnetization versus applied magnetic field $M(H)$, magnetic susceptibility $\chi(T)$ and heat capacity $C(T)$ measurements of BaMn$_2$As$_2$ single crystals.

\section{Experimental DETAILS}
\label{sec:EXPT}
Single crystals of BaMn$_2$As$_2$ were grown out of MnAs and Sn fluxes.  For the growth with Sn flux the elements were taken in the ratio Ba:Mn:As:Sn~=~ 1~:~2~:~2~:~35, placed in an alumina crucible and then sealed in a quartz tube under vacuum ($\approx$~10$^{-2}$~mbar).  The whole assembly was placed in a box furnace and heated to 1000~$^\circ$C at a rate of 50~$^\circ$C/hr, left there for 10~hrs and then cooled to 500~$^\circ$C at a rate of 5~$^\circ$C/hr.  At this temperature the molten Sn flux was decanted using a centrifuge.  Shiny plate-like crystals of typical size $2 \times 2 \times 0.1$~mm$^3$ were obtained.  

For crystal growth using MnAs flux small pieces of Ba metal and prereacted MnAs powder were taken in the ratio Ba:MnAs~=~1~:~5, placed in an alumina crucible and sealed in a quartz tube under a partial pressure of argon.  The whole assembly was placed in a box furnace and heated to 1180~$^\circ$C at a rate of 50~$^\circ$C/hr, left there for 6~hrs and then cooled to 1050~$^\circ$C at a rate of 5~$^\circ$C/hr.  At this temperature the excess MnAs flux was decanted using a centrifuge.  Plate-like crystals of typical size $ 2.5 \times 2.5 \times 0.2$~mm$^3$ were obtained.  

Crystals grown from both fluxes were extremely malleable and could be easily bent.  The compositions of two crystals, one from each type of growth, were checked using energy dispersive x-ray (EDX) semiquantitative analysis using a JEOL scanning electron microscope (SEM).  The SEM scans were taken on cleaved surfaces of the crystals.  For the crystal grown out of Sn flux the EDX gave the average elemental ratio Ba:Mn:As:Sn~=~19.6~:~41.5~:~38.8:~0.1 which is consistent with an approximate 1:2:2 stoichiometry for the compound and almost no Sn inclusion in the crystals.  The Sn concentration error is consistent with zero Sn content.  For the crystal grown from MnAs flux the EDX gave the average elemental ratio Ba:Mn:As~=~20.8~:~41.2~:~38.  The EDX measurements on the crystals did not show the presence of any other elements.  Laue x-ray back-scattering measurements on the crystals showed that the largest surface of the plates was perpendicular to the $c$ axis. 

For crystal structure determination, single crystal x-ray diffraction measurements on a Sn flux-grown single crystal were done at temperature $T$~=~293~K using a Bruker CCD-1000 diffractometer with Mo $K_{\alpha}$ ($\lambda$~=~0.71073~\AA) radiation.  Powder X-ray diffraction (XRD) measurements were done on crushed crystals of BaMn$_2$As$_2$ grown out of MnAs flux.  The XRD patterns were obtained at room temperature using a Rigaku Geigerflex diffractometer with Cu K$\alpha$ radiation, in the 2$\theta$ range from 10 to 90$^\circ$ with a 0.02$^\circ$ step size. Intensity data were accumulated for 5~s per step. 
The anisotropic magnetic susceptibility $\chi$ versus temperature $T$ and magnetization $M$ versus magnetic field $H$ measurements were done using a commercial Quantum Design SQUID magnetometer on a 1.65~mg single crystal grown out of Sn flux.  The standard four-probe $\rho(T)$ was measured with a current of amplitude $I$~=~1~mA at a frequency of 16~Hz, using the ac transport option of a commercial Physical Property Measurement System (PPMS, Quantum Design).  The contacts were made with silver epoxy on a cleaved surface of a crystal.  The current was applied in the $ab$ plane.  The $C(T)$ was measured on a MnAs-grown single crystal of mass 5.8~mg using the commercial PPMS.

\section{RESULTS}
\subsection{Single Crystal Structure Determination and Powder X-ray Diffraction of BaMn$_2$As$_2$}
\label{sec:RES-structure}
Our x-ray diffraction measurements revealed no impurity phases in the crystals.  A $0.22 \times 0.2 \times 0.03$~mm$^3$ plate-like single crystal grown out of Sn flux was used for single crystal structure determination.  The initial cell parameters were taken as those previously reported for polycrystalline BaMn$_2$As$_2$ (ThCr$_2$Si$_2$ structure, $Z$~=~2 formula units/unit cell, space group $I4/mmm$).\cite{Brechtel1978}  The final cell parameters and atomic positions were calculated from a set of 873 strong reflections with good profiles in the range 2$\theta$~=~6$^\circ$--61$^\circ$.  The unit cell parameters were found to be $a$~=~$b$~=~4.1686(4)~\AA~ and $c$~=~13.473(3)~\AA~ for the Sn-grown crystals.  A Rietveld refinement\cite{Rietveld} of the powder XRD pattern of MnAs-grown BaMn$_2$As$_2$ gave the lattice parameters $a$~=~$b$~=~4.1672(3)~\AA~ and $c$~=~13.466(5)~\AA.  These values are in good agreement with previously reported values for polycrystalline BaMn$_2$As$_2$ [$a$~=~$b$~=~4.15~\AA, and $c$~=~13.47 \AA ].\cite{Brechtel1978}  There is only one atomic coordinate not constrained by symmetry requirements, the $z$ position for As.  We find $z$~=~0.3615(3) for both MnAs- and Sn-grown crystals.  The single crystals of BaMn$_2$As$_2$ have a high tendency to split/cleave into very thin plates perpendicular to the $c$ axis resulting in significant mosaicity/twinning.  This resulted in some broadened reflections and did not allow a full refinement of the structure with a reasonable ($< 10\%$) reliability factor.

\subsection{Magnetization and Magnetic Susceptibility}
\label{sec:MAG}
\begin{figure}[t]
\includegraphics[width=3in]{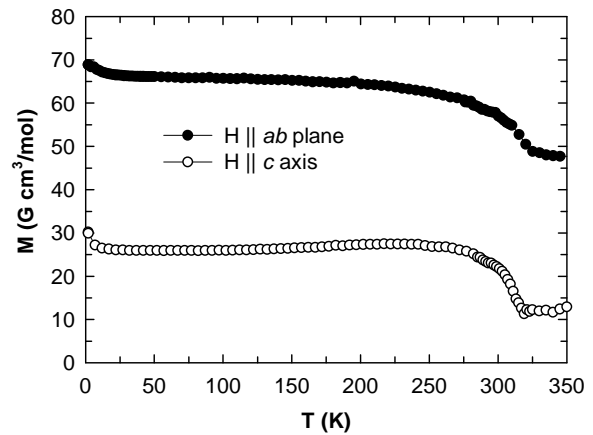}
\caption{The observed magnetization $M$ versus temperature $T$ in a magnetic field $H$~=~3~T applied in the $ab$ plane and along the $c$ axis for a crystal of BaMn$_2$As$_2$ grown from Sn flux.
\label{FigM(T)}}
\end{figure}

The magnetization $M$ versus temperature $T$ measured in a magnetic field $H$~=~3~T applied in the $ab$ plane $M_{ab}(T)$ and with $H$ along the $c$ axis $M_{c}(T)$ for a single crystal of BaMn$_2$As$_2$ of mass $m$~=~0.340~mg grown out of Sn flux is shown in Fig.~\ref{FigM(T)}.  The sudden increase in the $M(T)$ data seen upon cooling below about $T\approx 310$~K for both directions most likely arises from a small amount of MnAs impurity which is known to undergo a first order ferromagnetic/structural transition around $T\approx$~318~K\@.\cite{Bean1962}  

\begin{figure}[t]
\includegraphics[width=3in]{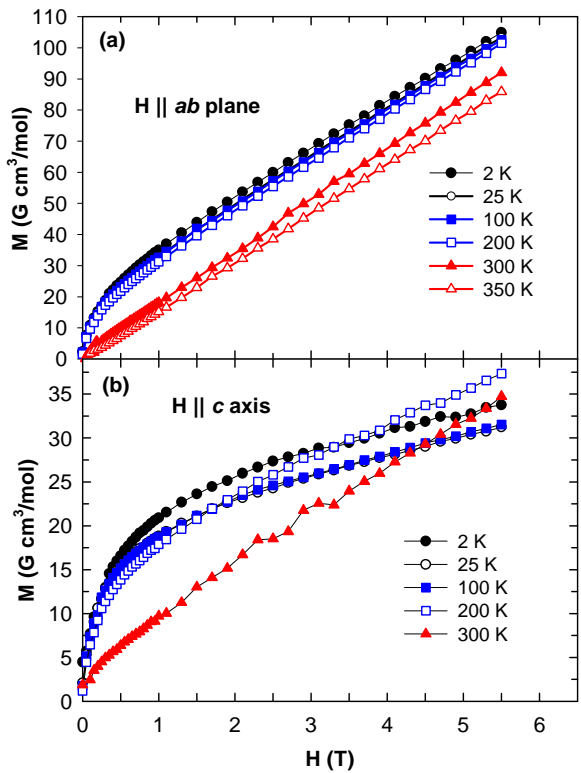}
\caption{ Magnetization $M$ versus magnetic field $H$ at various temperatures $T$ with (a) $H$ in the $ab$ plane and with (b) $H$ along the $c$ axis for a crystal of BaMn$_2$As$_2$ grown from Sn flux.
\label{FigM(H)}}
\end{figure}

\begin{figure}[t]
\includegraphics[width=3in]{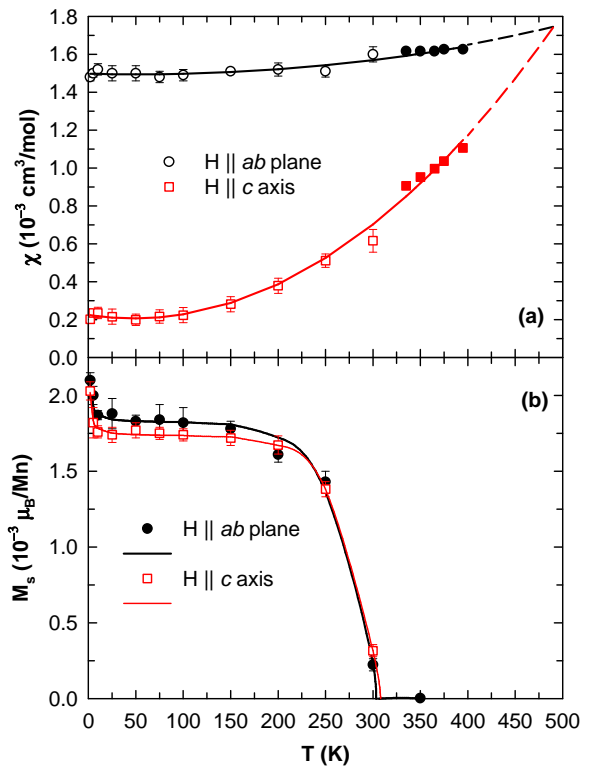}
\caption{(a) Magnetic susceptibilities $\chi_{ab}$ and $\chi_{c}$ versus temperature $T$ for a crystal of BaMn$_2$As$_2$ grown from Sn flux.  The open symbols are the data extracted from $M(H)$ isotherms and the closed symbols are the $\chi \equiv M/H$ data obtained from $M(T)$ measurements at fixed $H$~=~3~T\@.  The solid curves through the $\chi_{ab}(T)$ and $\chi_{c}(T)$ data are fits by second order polynomials.   The fits have been extrapolated (dashed curves) to higher $T$ where they intersect at the extrapolated antiferromagnetic ordering temperature $T_{\rm N}\sim$~500~K\@. (b) The saturation magnetization $M_{\rm s}$ versus $T$ for $H$ in the $ab$ plane and along the $c$ axis for a crystal of BaMn$_2$As$_2$ grown from Sn flux.  The solid curves through the data are guides to the eye.  
\label{Figchi}}
\end{figure}

To extract the intrinsic magnetic behavior of BaMn$_2$As$_2$ we have carried out $M(H)$ isotherm measurements.  Representative $M(H)$ isotherm data are shown in Figs.~\ref{FigM(H)}(a) and~(b) for $H$ in the $ab$ plane and for $H$ along the $c$ axis, respectively.  The $M(H)$ data for $T$~=~300~K and 350~K are linear (except at low $H$).  The $M(H)$ data for lower $T$ show a rapid increase at low $H$, with a tendency of saturating around $H$~=~1--2~T, and then show a linear behavior for higher $H$.  These data indicate that the contribution from the MnAs impurity saturates by about $H$~=~2~T\@.  Linear fits to the $M(H)$ data for the field range $H$~=~3--5.5~T were performed for the $M(H)$ data.  The slope of the high-field linear fits at various $T$ gave the intrinsic susceptibility $\chi(T)$ of BaMn$_2$As$_2$ and the $H$~=~0~T intercept gave the saturation magnetization $M_{\rm s}$ versus $T$ which are shown in Figs.~\ref{Figchi}(a) and~(b), respectively.\cite{MnAs}  Above $T \simeq$ 25~K the $M_{\rm s}(T)$ is attributed to the saturation magnetization of the MnAs ferromagnetic impurity phase.  The small upturn at lower temperatures is likely due to saturation of paramagnetic impurities.  From Fig.~\ref{Figchi}(b), the $M_{\rm s}$ is seen to be negligible above about 320~K\@.  Also from Fig.~\ref{Figchi}(b) it can be seen that $M_{\rm s}$ for $H$ along the $ab$ plane and $H$ along the $c$ axis have the same $T$ dependence and are nearly isotropic as expected.  The value of $M_{\rm s}$~=~3.6$\times 10^{-3}~\mu_{\rm B}$/f.u. (f.u. means formula unit) at $T$~=~25~K indicates a very small concentration of about 0.11~mol\% MnAs impurities [$M_{\rm s} = 3.40(3)~\mu_{\rm B}$/Mn for MnAs at $T$~=~0~K].\cite{Haneda1977}   

The $\chi(T)$ data obtained up to 350~K from the $M(H)$ isotherms are shown as open symbols in Fig.~\ref{Figchi}(a).  We carried out additional measurements of $\chi \equiv M/H$ in $H$~=~3~T at higher temperatures up to 400~K, shown as filled symbols in Fig.~\ref{Figchi}(a).  The intrinsic $\chi(T)$ data extracted from the $M(H)$ data and the additional $\chi(T)$ data match very well for both field directions.  We find $\chi_{ab}$/$\chi_{c}$~=~7.5 at $T$~=~2~K\@.  The in-plane susceptibility $\chi_{ab}\approx$~1.5$\times 10^{-3}$~cm$^3$/mol is nearly $T$ independent in the temperature range of the measurements, whereas the $c$ axis susceptibility $\chi_{c}\approx$~2$\times 10^{-4}$~cm$^3$/mol is nearly $T$ independent between $T$~=~2~K and $T$~=~100~K and then increases with increasing $T$ up to our maximum measurement temperature of 400~K\@.  

The anisotropic $\chi(T)$ data in Fig.~\ref{Figchi}(a) strongly suggest that BaMn$_2$As$_2$ is antiferromagnetically ordered at room temperature and below.  The anisotropy is the same as expected for a fiducial mean-field collinear antiferromagnet, where the spin susceptibility along the easy-axis direction goes to zero for $T \to 0$, and the spin susceptibility perpendicular to the easy-axis direction is nearly constant below the N\'eel temperature.  Thus from Fig.~\ref{Figchi}(a) we identify the easy-axis to be the $c$ axis.  The reason for the small positive value of  $\chi_{c}(T\to 0)$ is probably due to the presence of a paramagnetic orbital (Van Vleck) susceptibility.  Extrapolation of the two data sets in Fig.~\ref{Figchi}(a) to higher temperatures [dashed curves in Fig.~\ref{Figchi}(a)] suggests that $T_{\rm N} \sim$~500~K\@.  High antiferromagnetic ordering temperatures ($\sim$~400~K) have been observed in the isostructural Mn compounds $A$Mn$_2$Ge$_2$ ($A$~=~Ca and Ba) \cite{Malaman1994} and BaMn$_2$P$_2$ ($T_{\rm N} >$~750~K).\cite{Brock1994}  

\subsection{Resistivity}
\label{sec:RES-resistivity}
\begin{figure}[t]
\includegraphics[width=3in]{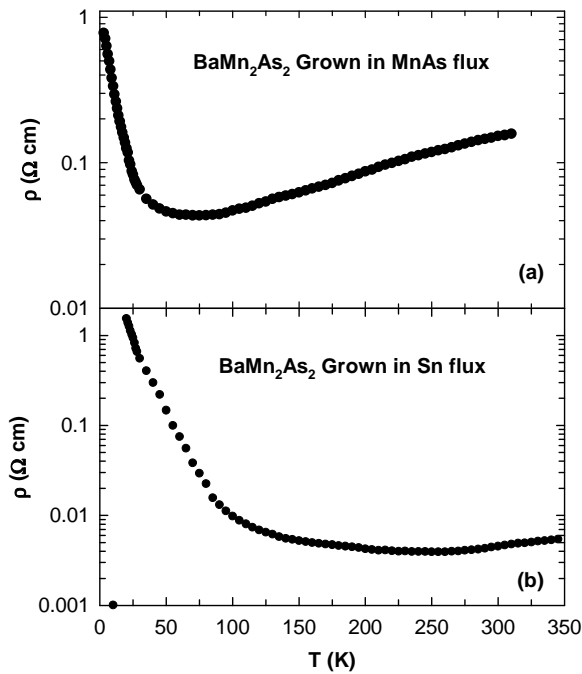}
\caption{Resistivity $\rho$ on a log scale versus temperature $T$ in the $ab$ plane for a crystal of BaMn$_2$As$_2$ grown (a) from MnAs flux and (b) from Sn flux.
\label{Figres}}
\end{figure}

The electrical resistivity $\rho$ versus temperature $T$ in the $ab$ plane for the BaMn$_2$As$_2$ crystals grown from MnAs flux and from Sn flux are shown in Figs.~\ref{Figres}(a) and \ref{Figres}(b), respectively.  For the MnAs-grown crystal the $\rho(T)$ data show a monotonic decrease with decreasing $T$ between 310~K and 70~K before increasing for lower $T$.  For the Sn-grown crystal $\rho$ decreases with decreasing $T$ between $T$~=~350~K and 250~K, stays almost $T$ independent between $T$~=~250~K and $T$~=~100~K, and increases strongly for lower $T$.  The $T$ dependence of $\rho$ for crystals grown from MnAs and Sn fluxes are qualitatively similar.  The difference could occur from a small inclusion of Sn flux in the BaMn$_2$As$_2$ crystals grown out of Sn and/or from slightly different compositions and/or defect concentrations of the crystals.   

\begin{figure}[t]
\includegraphics[width=3in]{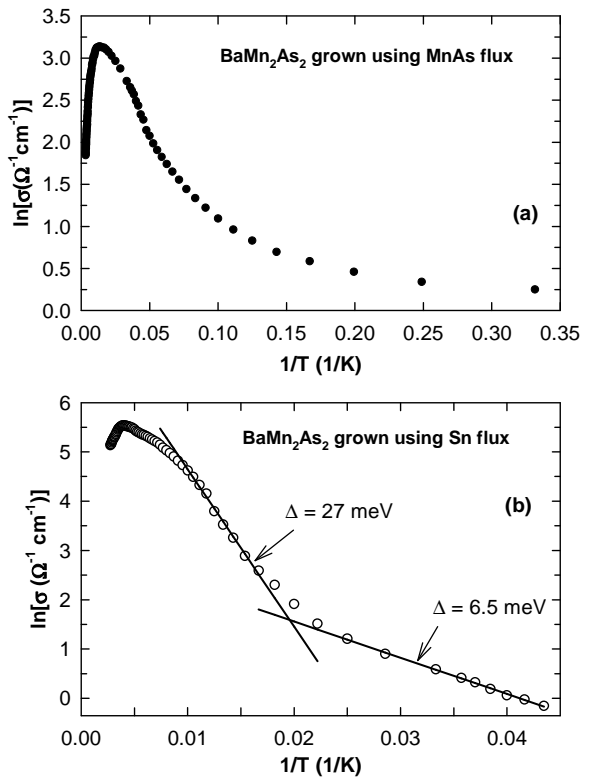}
\caption{ln($\sigma$) versus inverse temperature $1/T$ (where conductivity $\sigma = 1/\rho$ from Fig.~\ref{Figres}) for a crystal of BaMn$_2$As$_2$ grown (a) from MnAs flux and (b) from Sn flux.  The solid curves through the data in (b) are fits over restricted temperature intervals by the expression ln($\sigma$)~=~$A - \Delta/T$.  
\label{Figresfit}}
\end{figure}

The ln($\sigma$) versus $1/T$ data (where conductivity $\sigma = 1/\rho$ from Fig.~\ref{Figres}) are shown in Figs.~\ref{Figresfit}(a) and \ref{Figresfit}(b) for MnAs-grown and Sn-grown BaMn$_2$As$_2$ crystals, respectively.  For the MnAs-grown crystal, no extended linear regions in $T$ were found.  For the Sn-grown crystal the data between $T$~=~60~K and 100~K and between $T$~=~20~K and 40~K were found to be nearly linear in $T$ and were fitted by the expression ln($\sigma$)~=~$A - \Delta/T$, where $A$ is a constant and $\Delta$ is the activation energy.  The fits, shown as the solid curves through the data in Fig.~\ref{Figresfit} gave the values $\Delta$~=~27~meV and 6.5~meV for the two fits as shown with arrows in Fig.~\ref{Figresfit}(b).  

Our resistivity $\rho(T)$ and conductivity $\sigma(T) = 1/\rho(T)$ data for BaMn$_2$As$_2$ in Figs.~\ref{Figres} and~\ref{Figresfit}, respectively, indicate that this compound is a doped semiconductor as follows.  The conductivity can be written as $\sigma = n_{\rm e}e\mu_{\rm e} + n_{\rm h}e\mu_{\rm h}$, where $n_{\rm e} (n_{\rm h})$ is the concentration of electrons (holes) and $\mu_{\rm e} (\mu_{\rm h})$ is the electron (hole) mobility.\cite{Blatt1968}  In a semiconductor, the carrier concentration increases and the carrier mobility generally decreases with increasing $T$.  We infer that the ``metallic'' behavior of $\rho(T)$ at high temperatures in Fig.~\ref{Figres}, defined as where $\rho(T)$ has a positive temperature coefficient, could occur because the mobility of the carriers decreases faster than the carrier concentration increases with increasing $T$.\cite{Morin1954}  At lower temperatures, the situation is reversed and the resistivity increases with decreasing $T$.  The existence of two distinct slopes in Fig.~\ref{Figresfit}(b) suggests that the larger slope is the intrinsic activation energy (which is one-half the energy gap between valence and conduction bands if they have the same magnitude of curvature at the band edges) and the smaller value is the energy gap between donor or acceptor energy levels and the conduction or valence band, respectively.\cite{Pearson1949}  Thus we infer that at least at low temperatures, BaMn$_2$As$_2$ is a small band gap semiconductor with an intrinsic activation energy of order 0.03~eV and with an insulating ground state.  This activation energy is of the same order as previously found (0.07~eV) for BaMn$_2$P$_2$.\cite{Brock1994}

We consider the following alternative model for the resistivity at high temperatures.  The value of the resistivity at its minimum is $\rho \approx 44$~m$\Omega$~cm at $T = 75$~K for the MnAs-grown crystal and is $\rho \approx  4$~m$\Omega$-cm at $T = 250$~K for the Sn-grown crystal.  Such high values of $\rho$ for a metal and the positive temperature coefficient of $\rho(T)$ at high temperatures are characteristic of a so-called ``bad metal'',\cite{Hussey2004} where the mean free path for conduction carrier scattering is of order or less than an interatomic distance.  In this scenario, a metal-insulator (-semiconductor) transition or crossover would evidently occur on cooling into the temperature region of semiconductor-like resistivity behavior.

\subsection{Heat Capacity}
\label{sec:RES-heatcapacity}
\begin{figure}[t]
\includegraphics[width=3in]{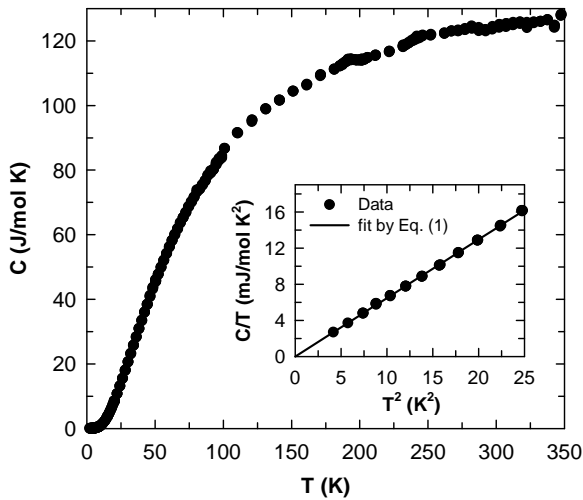}
\caption{Heat capacity $C$ versus temperature $T$ for a MnAs-grown single crystal of BaMn$_2$As$_2$ between 2~K and 300~K\@.  The inset shows the $C(T)/T$ versus $T^2$ data below $T$~=~5~K\@.  The straight line through the data in the inset is a fit by the expression $C/T$~=~$\gamma$+$\beta T^2$.  
\label{FigHC}}
\end{figure}

The heat capacity $C$ versus temperature $T$ between 2~K and 300~K of a BaMn$_2$As$_2$ crystal grown from MnAs flux is shown in Fig.~\ref{FigHC}.  The heat capacity at room temperature $C(300~{\rm K})\approx$~130~J/mol~K is close to the classical Dulong Petit lattice heat capacity value $C$~=~15$R\approx$~125~J/mol~K expected for BaMn$_2$As$_2$, where $R$ is the molar gas constant.  There is no clear signature of any phase transition in the temperature range of our measurements.  However, two small bumps at $T \sim$~180~K and 230~K are seen in the $C(T)$ data although these are probably noise since there are no corresponding anomalies in the $\chi(T)$ or $\rho(T)$ data at these temperatures.  

\noindent

Figure~\ref{FigHC} inset shows the $C(T)/T$ versus $T^2$ data between 2~K and 5~K\@.  The $C(T)/T$ data are linear in $T^2$ in this temperature range and were fitted by the expression 
\begin{equation}
C/T = \gamma + \beta T^2~, 
\label{Eqgamma}
\end{equation}
\noindent
where $\gamma$ is the Sommerfeld coefficient of the electronic heat capacity.  We obtain $\gamma$~=~0.0(4)~mJ/mol~K$^2$ and $\beta$~=~0.65(3)~mJ/mol~K$^4$.  The value of $\gamma$ is consistent with the value zero and confirms the nonmetallic ground state indicated by the transport measurements above.   

If we assume that the $\beta T^2$ term in Eq.~(\ref{Eqgamma}) arises from the lattice heat capacity, from the value of $\beta$ estimated above one can obtain the Debye temperature $\Theta_{\rm D}$ using the expression \cite{Kittel}
\begin{equation}
\Theta_{\rm D}~=~\bigg({12\pi^4Rn \over 5\beta}\bigg)^{1/3}~, 
\label{EqDebyetemp}
\end{equation}
\noindent
where $n$ is the number of atoms per formula unit (\emph{n}~=~5 for BaRh$_2$As$_2$).  We obtain $\Theta_{\rm D}$~=~246(4)~K for BaMn$_2$As$_2$.  However, since BaMn$_2$As$_2$ is probably antiferromagnetically ordered at low $T$, a small contribution to $\beta$ could be present due to excitations of antiferromagnetic spin waves.\cite{gopal} 

\section{CONCLUSION}
\label{sec:CON}
We have synthesized single crystalline samples of the layered arsenide BaMn$_2$As$_2$ using MnAs and Sn fluxes.  Single crystal structure determination and x-ray powder diffraction confirm that BaMn$_2$As$_2$ crystallizes in the tetragonal ThCr$_2$Si$_2$-type structure with lattice parameters $a$~=~$b$~=~4.1642(3)~\AA~ and $c$~=~13.454(5)~\AA~ for the MnAs-grown crystals and $a$~=~$b$~=~4.1686(4)~\AA~ and $c$~=~13.473(3)~\AA~ for the Sn-grown crystals.  

Electrical resistivity $\rho$ versus $T$ measurements above $T$~=~70~K for the MnAs-grown crystals, and above $T$~=~100~K for the Sn-grown crystals, show a metallic-like behavior with a decrease in $\rho$ with decreasing $T$.  On reducing $T$, $\rho$ reaches a minimum and then increases with decreasing $T$ for both kinds of crystals.  We estimate a small intrinsic activation energy of order 0.03~eV below the $T$ at which the minimum occurs.  The heat capacity versus $T$ measurements between 2~K and 5~K give a Sommerfeld coefficient of the linear heat capacity $\gamma$~=~0.0(4)~mJ/mol~K$^2$ and a Debye temperature $\theta_{\rm D}$~=~246(4)~K\@.  The value of $\gamma$ is consistent with zero and suggests almost zero density of states at the Fermi energy.  Thus the results of our transport and thermal measurements consistently indicate that BaMn$_2$As$_2$ is a small band-gap semiconductor with an insulating ground state.  At high temperatures, since the resistivity has a positive temperature coefficient and a magnitude similar to that expected for a ``bad metal'',\cite{Hussey2004} it is possible that the material is a bad metal at high $T$ and exhibits a metal to insulator (semiconductor) transition or crossover with decreasing $T$.
 
Magnetization versus field and temperature measurements reveal an anisotropic magnetic susceptibility with $\chi_{ab}$/$\chi_c$~=~7.5 at $T$~=~2~K\@.  The $\chi_{ab}\approx$~1.5$\times 10^{-3}$~cm$^3$/mol is nearly $T$ independent in the temperature range of the measurements whereas $\chi_{c}\approx$~0.2$\times 10^{-3}$~cm$^3$/mol is nearly $T$ independent between $T$~=~2~K and $T$~=~100~K and then increases with increasing $T$ up to $T$~=~400~K\@.  These data suggest that BaMn$_2$As$_2$ is antiferromagnetically ordered at room temperature, with the $c$ axis being the easy axis, and with an extrapolated N\'eel temperature $T_{\rm N}\sim$~500~K\@.  If this is confirmed, this would be an interesting result because the magnetic order occurs without a concomitant crystallographic phase transition, in contrast to the (Ca,Sr,Ba)Fe$_2$As$_2$ compounds.\cite{Sadovskii2008}    

It is of interest to compare and contrast the properties of BaMn$_2$As$_2$ with those of (Ca,Sr,Ba)Fe$_2$As$_2$ (Ref.~\onlinecite{Sadovskii2008}) and the layered cuprate compounds like La$_2$CuO$_4$,\cite{Johnston1997} the latter two of which are both known ``parent compounds'' for high temperature superconductors.  As we have shown, BaMn$_2$As$_2$ has an insulating ground state like La$_2$CuO$_4$ but with a much smaller energy gap, whereas the (Ca,Sr,Ba)Fe$_2$As$_2$ compounds are metals.  BaMn$_2$As$_2$ (as we infer) and La$_2$CuO$_4$ are both local moment antiferromagnetic insulators, whereas the (Ca,Sr,Ba)Fe$_2$As$_2$ compounds are widely regarded as correlated itinerant spin-density-wave (SDW) materials.  Our extrapolated N\'eel temperature $T_{\rm N} \sim 500$~K for BaMn$_2$As$_2$ is high compared to the SDW transition temperatures $\lesssim 200$~K for the (Ca,Sr,Ba)Fe$_2$As$_2$ materials.  The $T_{\rm N}$ of La$_2$CuO$_4$ is 325~K, but this is strongly suppressed from the mean field value $\sim 1600$~K by fluctuation effects associated with the two-dimensionality of the Cu spin lattice.\cite{Johnston1997}  Thus with respect to ionicity/covalency, it appears that BaMn$_2$As$_2$ is intermediate between (Ca,Sr,Ba)Fe$_2$As$_2$ and the layered cuprate compounds and thus forms a bridge between these other two classes of materials.  Theoretically, one expects a maximum in the superconducting transition temperature $T_{\rm c}$ as a function of superconducting pair electronic coupling strength.\cite{Chubukov2005}  If we associate this coupling strength with the degree of ionicity/covalency in a material,\cite{Schmalian2008} this suggests that the parent compound BaMn$_2$As$_2$ could be closer to this maximum than either of the other two types of materials, and that doping or pressurizing BaMn$_2$As$_2$ might possibly lead to exceptionally high values of $T_{\rm c}$.

\noindent
\emph{Note added}--- While we were writing this paper, a preprint appeared \cite{An2009} on the band structure calculation and electrical resistivity and heat capacity measurements on single crystals of BaMn$_2$As$_2$ grown using MnAs flux.  Our resistivity and heat capacity measurement results are qualitatively similar to those reported in this preprint.  

\begin{acknowledgments}
We are grateful to J. Schmalian for helpful discussions.   Work at the Ames Laboratory was supported by the Department of Energy-Basic Energy Sciences under Contract No.\ DE-AC02-07CH11358.  
\end{acknowledgments}


\begin{references}
\bibitem{MnAs} The MnAs-grown crystals showed $\sim 5$ times larger magnetization contribution from MnAs impurity at 2~K and were therefore not used to extract the intrinsic susceptibility of BaMn$_2$As$_2$.
\end{references}
\end{document}